\documentclass[Transaction]{IEEEtran}
\usepackage{amsfonts}
\usepackage{amssymb}
\usepackage{amsmath}
\usepackage{graphicx}
\usepackage{subfigure}
\usepackage{cite}
\usepackage{setspace}
\usepackage{algorithm}
\usepackage{algorithmic}
\usepackage{multirow}
\usepackage{amsmath}
\usepackage{xcolor}
\usepackage{mathrsfs}
\usepackage{dsfont}
\usepackage{bm}
\usepackage{multicol}
\usepackage{amsfonts,amssymb}
\usepackage{array,color}
\usepackage{bbm}

\begin{document}

\title{Inter-tier Interference Suppression in Heterogeneous Cloud Radio Access
Networks
\thanks{The work of Mugen Peng was supported in part by the National Natural Science Foundation
of China (Grant No.61361166005), the National Basic Research Program
of China (973 Program) (Grant No. 2013CB336600), and the Beijing Natural Science Foundation (Grant No.
4131003). The work of H. Vincent Poor was supported in part by the U.S. National Science Foundation under Grant ECCS-1343210.}
\thanks{Mugen Peng, Hongyu Xiang, Yuanyuan Cheng, and Shi Yan are with the Key Laboratory of Universal Wireless Communications for Ministry of Education, Beijing University of Posts and Telecommunications, Beijing, China (e-mail: pmg@bupt.edu.cn; xianghongyu88@163.com; chinachengyuanyuan@gmail.com; yanshi01@bupt.edu.cn). }
\thanks{H. Vincent Poor is with the School of
Engineering and Applied Science, Princeton University, Princeton,
NJ, USA (e-mail: poor@princeton.edu). }}
\author{\IEEEauthorblockN{Mugen Peng, Hongyu Xiang, Yuanyuan Cheng, Shi Yan, and H. Vincent Poor}}

\maketitle

\begin{abstract}

Incorporating cloud computing into heterogeneous networks, the
heterogeneous cloud radio access network (H-CRAN) has been proposed as a
promising paradigm to enhance both spectral and energy efficiencies.
Developing interference suppression strategies is critical for suppressing the
inter-tier interference between remote radio heads (RRHs) and a macro
base station (MBS) in H-CRANs. In this paper, inter-tier interference suppression techniques are considered in the contexts of collaborative processing and cooperative radio resource allocation (CRRA). In particular, interference collaboration (IC) and beamforming (BF) are proposed to suppress the inter-tier interference, and their corresponding performance is evaluated. Closed-form expressions for the overall outage probabilities, system capacities, and average bit error rates under these two schemes are derived. Furthermore, IC and BF based CRRA optimization models are presented
to maximize the RRH-accessed users' sum rates via power allocation, which is solved with convex optimization. Simulation results demonstrate that the derived expressions for these performance metrics for IC and BF are accurate; and the relative performance between IC and BF schemes depends on system parameters, such as the number of antennas at the MBS, the number of RRHs, and the target signal-to-interference-plus-noise ratio threshold. Furthermore,
it is seen that the sum rates of IC and BF schemes increase almost linearly with the transmit power threshold
under the proposed CRRA optimization solution.

\end{abstract}


\section{Introduction}
To meet the rapidly growing mobile data volume driven by applications on platforms such as smartphones and tablets,
the next generation of wireless networks face significant challenges in improving system capacity and guaranteeing users' quality of service (QoS)~\cite{data}. Cloud radio access networks (C-RANs)
have been proposed to provide high bit rates, while reducing
both capital and operating expenditures~\cite{CRAN1}~\cite{CRAN2}. By migrating the baseband
functionalities of base stations (BSs) to a centralized baseband unit (BBU) pool and distributed remote radio heads (RRHs), the C-RAN facilitates the implementation of centralized coordinated multi point (CoMP) transmission~\cite{CoMP}. With such an architecture, mobile operators can easily expand and upgrade their networks
by deploying additional RRHs, and thus the
corresponding operational costs can be greatly reduced. Unfortunately, one of the main restrictions on the implementation of C-RANs is the non-ideal fronthaul
with limited capacity and long time delay. Overcoming
the negative impact of the constrained fronthaul on spectral efficiency (SE) and energy efficiency (EE) is not straightforward~\cite{EE}.

The heterogeneous cloud radio access network (H-CRAN) has recently
been proposed to decouple the control plane and user plane to
enhance the existing C-RAN concept, in which the functions of
control plane are only implemented in traditional macro base stations
(MBSs)~\cite{HCRAN}. In H-CRANs, RRHs are used to provide high bit rates for users with diverse QoS requirements
in hot spots, while the MBS is deployed to guarantee seamless coverage and deliver the
control signalling of the whole network. User equipments (UEs) can access RRHs transparently in H-CRANs, which allows UEs to operate over a single carrier frequency and at low cost.
In comparison with C-RANs and heterogenous networks (HetNets)~\cite{hetnet}, H-CRANs have been demonstrated to achieve significant performance gains through advanced collaborative signal processing. However, because MBSs and RRHs are underlaid with the same carrier frequency in the same coverage area, severe inter-tier interference is incurred, which degrades the performance of H-CRANs significantly.

Unlike the traditional HetNets, the intra-tier interference among dense RRHs in H-CRANs can be fully eliminated by large-scale cooperative processing through the fronthaul, while the inter-cell interference between adjacent BSs in HetNets should be mitigated by the distributed CoMP techniques through backhaul. Furthermore, the inter-tier interference to the RRH user equipments (RUEs) in H-CRANs can be coordinated through spatial multiple-input and multiple-output (MIMO) processing in the MBS with multiple antennas~\cite{MIMO}, and the inter-tier interference at the MBS user equipments (MUEs) can be endured because the MUE provides seamless coverage at only low bit rates. As a result, the inter-tier interference to RUEs makes it challenging to improve SE in H-CRANs, and thus advanced inter-tier interference suppression techniques are of interest. Advanced inter-tier interference suppression techniques can be categorized as either collaborative processing in the physical layer or cooperative radio resource allocation (CRRA) in the upper layers~\cite{HetNet_peng}. This issue is the subject of this paper.

\subsection{Related Work}

Much recent attention has been paid to interference collaboration and CRRA for C-RANs. One of the key advantages of the C-RAN architecture is that it provides the BBU pool for joint baseband signal processing across the multiple RRHs in both uplink and downlink, and thus it achieves significantly higher data rates than conventional cellular networks. Thanks to the large-scale collaborative processing in the BBU, the intra-tier interference across RRHs can be fully eliminated. Such large-scale collaborative processing is often referred to as \textit{network precoding} or CoMP in HetNets.

Numerous studies of network precoding for MIMO systems, HetNets and C-RANs have been described in previous works~\cite{precoding}-\cite{andrews}. For example,
performance analysis under various linear precoding schemes has been presented in~\cite{precoding}, which can be directly applicable to H-CRANs.
The authors in~\cite{ZF} have analyzed the
throughput of multiuser MIMO for distributed antenna
systems based on zero-forcing beamforming (BF); however, the closed-form expressions for ergodic capacity therein have been presented with approximations
instead of exact results. In addition, in~\cite{CBF}, two coordinated BF designs have been taken into consideration in multicell networks: the QoS BF, and the max-min signal-to-interference-plus-noise ratio (SINR) BF. The goal of QoS BF is to minimize the total power consumption while guaranteeing that the received SINR of each user is above a pre-determined threshold, while the max-min SINR BF aims to maximize the minimum received SINR among all users under per-base-station power constraints.
Furthermore, there are two kinds of precoding schemes for MIMO, namely interference collaboration
(IC)~\cite{BFIC} and BF~\cite{BF}. In~\cite{BFIC},
an adaptive transmission strategy to switch between
IC and BF is proposed; however, the analytical results are restricted to the
scenario with only one low power node. Essentially, BF aims to maximize
the received signal strength for the desired users when the edge SINR is low, while IC is preferred when the
edge SINR is relatively high and the interference should be
suppressed~\cite{andrews}. Inspired by~\cite{ZF}-\cite{andrews}, in this paper both IC and BF schemes are extended to H-CRANs as advanced collaborative processing approaches to suppress the inter-tier interference, and the overall outage probability, system capacity, and average bit error rate (BER)
under IC and BF are used to evaluate their performance under different configurations.

In addition, to exploit the performance of C-RANs, the ergodic capacity performance of the single nearest and
\emph{N}-nearest association strategies with varying transmit power of RRHs in C-RANs is
compared in~\cite{CRAN}. The best RRH selection
scheme needs only a single RRH and hence reduces the system overhead by
avoiding coordination of the distributed RRHs, while resulting
in a certain performance loss. With the employment of precoding schemes, large-scale collaborative processing gains can be achieved in C-RANs with dense RRHs. It is indicated that no more than four RRHs should be associated for each UE to balance performance gains and implementation cost.
In~\cite{selection}, different performance metrics, such as outage probability, are used to compare downlink beamforming and
antenna selection, as well as
their impacts on reception reliability.

Besides the IC and BF schemes in the physical layer, inter-tier interference can be suppressed by cross-layer CRRA techniques in the upper layers. The significant cloud computing capability in the BBU pool enables the use of advanced cross-layer CRRA. Traditional radio resource allocation for cellular networks is largely based on heuristics and there is a lack of theoretical understanding of how to design cross-layer CRRA in an H-CRAN, which is usually more challenging than that in traditional cellular and C-RANs due to practical issues such as fronthaul capacity limitations, non-ideal channel state information (CSI), and the parallel implementation of algorithms.

Some optimization objectives like weighted
sum rate (WSR) for CRRA involve multi-user interference, causing non-convexity and making the problems hard to solve.
Fortunately, the weighted minimum mean square error (WMMSE) method has proven to be effective
in transforming such non-convex optimization problems into convex optimization problems.
Specifically, for the WSR problem with beamforming vectors as variables, the objective function
is non-convex with regard to the vectors. However, it has been shown that
WSR maximization and WMMSE minimization are equivalent for the MIMO interference channel \cite{mi}, in the sense that the two problems have the same optimal solution. Moreover, the obtained WMMSE minimization
after equivalent transformation is convex with respect to each of the individual optimization variables, and hence this non-convex problem is transformed into a more tractable convex
problem. As a result, the WMMSE method has been widely applied
to handle non-convex power consumption minimization \cite{1},
joint power and antenna selection optimization\cite{3}, and
weighted system throughput maximization \cite{IV:yu2015}.
Nevertheless, all of the above studies focus only on downlink transmission. In \cite{5}, the uplink transmission is taken into
consideration, and a joint downlink and uplink user-RRH association and precoding design scheme is proposed to minimize the system power consumption, in which the joint downlink and uplink optimization problem is transformed into an equivalent downlink problem, and the WMMSE method is used to transform the non-convex downlink problem into a convex problem with respect to the entries of the precoding matrix.

Moreover, the $l_0$-norm is often applied to express RRH selection, which leads to integer programming problems.
To transform such non-convex problems into convex problems, $l_1$-norm approximation can be used. In \cite{IV:yu2015}, the authors investigate re-weighted $l_1$-norm approximation in the fronthaul capacity constraint.
In the $l_1$-norm approximation method, each coefficient in the precoding matrix is assumed to be independent; however, such independence does not always hold in C-RANs. For example, one user is always served by a selected cluster of RRHs, which means that the elements not belonging to these RRHs in the precoding matrix are set to zero \cite{5}. Besides, one RRH can be switched off when all of its coefficients in the precoding matrix are set to zero \cite{IV:pre10}. In these cases, the coefficients of precoding matrices should be optimized jointly rather than individually, and thus the $l_1$-norm approximation cannot be used directly because the zero entries of the precoding matrices may not align in the same RRH.
To cope with this problem, the mixed $l_1/l_p$-norm approximation method can be adopted to induce group sparsity. In \cite{5} and \cite{IV:pre10}, mixed $l_1/l_p$-norm approximation methods are adopted to handle a group sparse based RRH selection problem. In \cite{5}, a traditional mixed $l_1/l_p$-norm method is used to transform group sparse based $l_0$-norm constraints. In \cite{IV:pre10}, a three stage group sparse precoding design algorithm is proposed to minimize the network energy consumption of C-RANs. The non-convex $l_0$-norm constraints are transformed into convex forms by a weighted mixed $l_1/l_p$-norm method. However, CRRA in H-CRANs for suppressing inter-tier interference when the precoding techniques with low complexity such as IC and BF are used in the physical layer has not been addressed.

\subsection{Contributions}

With the development of H-CRANs, the design of effective large-scale collaborative processing and cross-layer CRRA schemes for suppressing both intra-tier and inter-tier interference to improve SE is a key need. Considering the large-scale centralized collaborative processing in the BBU pool,
the intra-tier interference among RRHs can, in principle, be fully eliminated when the number of RRHs is not too large. Through the IC or BF based CRRA, the inter-tier interference in H-CRANs can be further suppressed. The major contributions of this paper can be summarized as follows.
\begin{itemize}
\item To mitigate the inter-tier interference between the MBS and RRHs in H-CRANs, IC and BF precoding schemes are employed at the multiple-antenna MBS. Performance metrics, including outage probability, system capacity, and average BER are analyzed for both IC and BF schemes. In particular, closed-form expressions for different performance metrics under IC and BF are derived.

\item Based on the derived closed-form expressions under IC and BF, the key factors, such as the number of antennas on the MBS, the number of RRHs, and the SINR threshold, impacting the overall outage probability, system capacity, and average BER are evaluated and compared.

\item Under the proposed IC and BF precoding schemes, CRRA to optimize RUEs' sum rates while guaranteeing the rates of MUEs is examined. The corresponding optimization problems based on both IC and BF are formulated as non-convex problems, which are solved by transforming them into convex problems and applying the Karush-Kuhn-Tucker (KKT) conditions. Based on the transformed Lagrangian function, the optimal power allocation algorithms for both RRHs and the MBS are developed.

\item The analytical and simulation results suggest that the IC and BF schemes should be adaptively switched between based on the system configuration and the adopted performance metrics. Meanwhile, the proposed CRRA solutions can achieve the optimal throughput by optimizing the transmit power. We see that the BF based CRRA outperforms the IC based CRRA in the regime of the high SINRs of the MBS, while the IC based CRRA outperforms the BF based CRRA in the regime of the low SINRs of the MBS.
\end{itemize}

The remainder of this paper is organized as follows. Section II
describes the H-CRAN system model and formulates the problem of interest. Section III analyzes the distribution of SINR for MUEs and RUEs under IC and BF precoding schemes. The outage probability, average BER, and system sum capacity under IC and BF schemes are derived in Section IV. Section V presents the BF and IC based CRRA optimization problems and the corresponding solutions. The simulation results for both collaborative processing in the physical layer and the CRRA in the upper layer are introduced in Section VI. Section VII summarizes this paper. For convenience, the
abbreviations are listed in Table I.

\begin{table}\label{table1}
\center \caption{Summary of Abbreviations}
\begin{tabular}{l l}\hline
CRRA & cooperative radio resource allocation\\
CSI & channel state information\\
EE & energy efficiency\\
HetNet & heterogenous network\\
H-CRAN & heterogeneous cloud radio access network\\
IC & interference collaboration\\
KKT & Karush-Kuhn-Tucker\\
MBS & macro base station\\
MIMO & multiple-input and multiple-output\\
MUE & MBS user equipment\\
QoS & quality of service\\
RRH & remote radio head\\
RUE & RRH user equipment\\
RV & random variable\\
MT & mobile terminal\\
SE & spectral efficiency\\
SINR & signal-to-interference-plus-noise ratio\\
UE & user equipment\\
WMMSE & weighted minimum mean square error\\
WSR & weighted sum rate\\
\hline
\end{tabular}
\end{table}

\section{H-CRAN System Model}

\begin{figure}[!h]
\centering
\includegraphics[width=7.5cm]{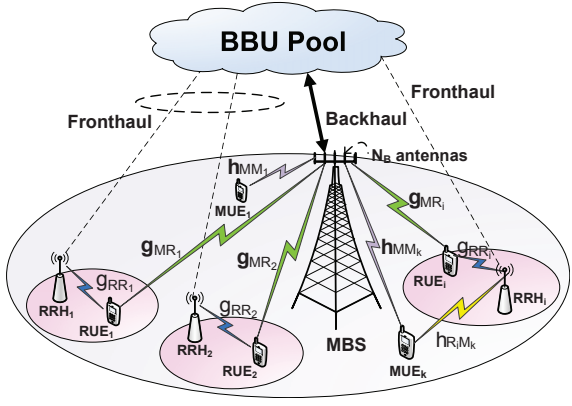}
\setlength{\belowcaptionskip}{-10pt}
\caption{System model of an H-CRAN with one MBS and $M$ RRHs}
\label{system}
\end{figure}

Unlike in C-RANs, the MBS in H-CRANs delivers the control signaling for the whole network, which decouples the user plane and control plane. Furthermore, to alleviate the heavy burdens on the fronthaul, some UEs with high mobility or with real-time traffic are given high priority to access the MBS. As a result, we can limit our attention to one MBS in the H-CRAN, under which multiple distributed RRHs are underlaid within the same coverage of the MBS. Thus, as illustrated in Fig. \ref{system}, the H-CRAN of interest consists
of one MBS and $M$ RRHs. For any typical radio resource block,
$K$ single-antenna MUEs
are served by the MBS, while only one single-antenna RUE is associated with each RRH. To serve multiple MUEs
simultaneously and suppress the inter-tier interference at RUEs in
the downlink, the MBS is equipped with $N_B$ antennas (${N_B} \ge M
+ K$), while each RRH is equipped with a single antenna.

The transmit power per antenna in the MBS and RRHs is assumed to be
$P_M$ and $P_R$, respectively. The transmission symbols for the
$j$-th MUE and the RUE associated with the $i$-th RRH are $s_{M_j}$ and $s_i$, respectively, which are normalized as ${\mathbb{E}}[{\left\| {{s_{M_j}}} \right\|^2}] = {\mathbb{E}}[{\left\| {{s_i}} \right\|^2}] = 1$. The received signal
at the $k$-th MUE and a typical RUE associated with the $i$-th RRH
can be written as
\begin{equation}\label{y}
\begin{gathered}
{y_{MM_{k}}}\! =\! \sum\limits_j^K{\sqrt {{P_M}}
{\textbf{h}_{MM_{k}}}\textbf{w}_j{s_{M_j}}}\!\! +\!\sum\limits_i^M
{\sqrt {P_{R}}} {h_{R_iM_{k}}}s_i\!\! +\! {n_{MM_{k}}},\hfill\\
{y_{RR_i}}\! =\! \sqrt {{P_R}} g_{R{R_i}}s_i \!+\! \sum\limits_j^K{\sqrt {{P_M}} {\textbf{g}_{M{R_i}}}\textbf{w}_j{s_{M_j}}} + n_{RR_i},\hfill\\
\end{gathered}
\end{equation}
respectively, where ${\textbf{h}_{MM_k}} \in {\mathbb{C}^{1 \times {N_B}}}$
represents the radio link between the MBS and the $k$-th MUE, and
${h_{{R_i}M_k}}$ represents the interference link from the
\emph{i}-th RRH to the $k$-th MUE. $\textbf{g}_{M{R_i}} \in {\mathbb{C}^{1
\times {N_B}}}$ represents the interference link between the MBS and the RUE
associated with the $i$-th RRH, and $g_{RR_i}$ represents the radio
link between the $i$-th RRH and its served RUE. Note that the inter-RRH interference amongst RRHs in H-CRANs can be ignored due to the centralized signal processing in the BBU pool through the ideal fronthaul.
We assume the radio links experience independent Rayleigh fading, so the
components of ${\textbf{h}_{MM_k}}$ and ${\textbf{g}_{M{R_i}}}$
are independent ${\cal C}{\cal N}(0, 1)$, ${h_{{R_i}M_k}} \sim {\cal
C}{\cal N}(0, 1)$, and ${g_{RR_{i}}} \sim {\cal C}{\cal N}(0, 1)$.
$n_{MM_{k}}$ and $n_{RR_{i}}$ are independent normalized additive zero-mean
Gaussian noises experienced at the $k$-th MUE and the
typical $i$-th RUE, respectively, i.e., $n_{MM_{k}}\sim {\cal C}{\cal N}(0, 1)$,
and $n_{RR_{i}} \sim {\cal C}{\cal N}(0, 1)$.
${\textbf{w}_{j}} \in {\mathbb{C}^{N_B \times 1}}$ represents the precoding
vector applied at the MBS for the $j$-th MUE.

According to (\ref{y}), the received SINR for the $k$-th MUE and the
typical RUE can be expressed as
\begin{equation}\label{MM}
{\gamma _{MM_k}} = \frac{{{P_M}{{\left|
{{\textbf{h}_{MM_{k}}}\textbf{w}_k} \right|}^2}}}{{\sum\limits_{j\!
=\! 1,j \ne k}^K {{P_M}{{\left| {{\textbf{h}_{MM_{k}}}\textbf{w}_j}
\right|}^2}}\! +\!\sum\limits_{i = 1}^M {{P_R}{{\left|
{{h_{{R_i}M_{k}}}} \right|}^2}} + 1 }},
\end{equation}
\begin{equation}\label{RR1}
{\gamma _{R{R_i}}} = \frac{{{P_R}{{\left| {{g_{R{R_i}}}}
\right|}^2}}}{{{P_M}\sum\limits_j^K{{\left|
{{\textbf{g}_{M{R_i}}}\textbf{w}_j} \right|}^2} + 1}},
\end{equation}
respectively. Since the interference is much larger than the noise
in an interference-limited H-CRAN, the noise could be ignored
herein. Thus (\ref{MM}) can be approximated as
\begin{equation}\label{start}
{\gamma _{MM_k}} \approx \frac{{{P_M}{{\left|
{{\textbf{h}_{MM_{k}}}\textbf{w}_k} \right|}^2}}}{{\sum\limits_{j =
1,j \ne k}^K {{P_M}{{\left| {{\textbf{h}_{MM_{k}}}\textbf{w}_j}
\right|}^2}}\! +\!\sum\limits_{i = 1}^M {{P_R}{{\left|
{{h_{{R_i}M_{k}}}} \right|}^2}}}}.
\end{equation}

The intra-tier interference among \emph{K} MUEs and
inter-tier interference between MUEs and RUEs can be suppressed by
precoding schemes in the MBS with multiple antennas.

\section{Inter-tier Collaborative Precoding schemes}

In this section, we describe two collaborative precoding schemes employed at the MBS with multiple antennas: IC and BF. The IC scheme enhances the performance gain by suppressing the interference to the RUEs and other MUEs, while BF based processing aims at maximizing the signal gain at the intended user and does not coordinate interference. We investigate the distribution of the SINR at the RUE and MUE, respectively, under these two precoding schemes in the following subsections.

\subsection{Interference Collaboration (IC)}

When the IC scheme is used at the MBS, the precoding vector
$\textbf{w}_k$ is chosen by nulling
the interference to the RUEs and other MUEs, which means $\textbf{w}_k
\in {\rm Null} (\tilde {\textbf{G}}_k)$, where $\tilde {\textbf{G}}_k=
[\textbf{g}_{{MR_1}}; \cdots;
\textbf{g}_{{MR_M}}; \textbf{h}_{{MM_1}}; \cdots; \textbf{h}_{{MM_{k-1}}};
\textbf{h}_{{MM_{k+1}}}; \cdots; \textbf{h}_{{MM_{K}}} ]$ $\in {{\mathbb{C}}^{(M + K - 1)\times N_B}}$, ${\rm Null}
(\tilde{\textbf{G}}_k) = \{ \textbf{v} \in {\mathbb{
C}^{{N_B} \times 1}}\!:\tilde {\textbf{G}}_k\textbf{v}=
\textbf{0}\}$, and ${\left(\cdot\right)}^H$
denotes the conjugate transpose. Furthermore, ${\textbf{g}_{MR_i}}\textbf{w}_k = 0$,
and ${\textbf{h}_{MM_j}}\textbf{w}_k = 0 \quad (\forall j \in K,j \ne k)$. Thus $\gamma _{RR_i}$ in (\ref{RR1}) and $\gamma
_{MM_k}$ in (\ref{start}) under IC can be simplified as
\begin{equation}\label{RR}
\begin{gathered}
{\gamma _{RR_i}^{IC}} = {P_R}{\left| {g_{R{R_i}}} \right|^2}, \hfill\\
{\gamma _{MM_k}^{IC}} = \frac{{{P_M}{{\left|
{{\textbf{h}_{MM_k}}\textbf{w}_k} \right|}^2}}}{{ \sum\nolimits_{i = 1}^M {{P_R}{{\left|{{h_{{R_i}M_k}}} \right|}^2}}}}. \hfill\\
\end{gathered}
\end{equation}

Hence from (\ref{RR}), ${\gamma _{RR_i}^{IC}} \sim {P_R}\chi
_{RR_{i}}^2 (2)$, where $\chi _{RR_{i}}^2 (2)$ denotes a chi-squared
random variable with two degrees of freedom. If the dimension of ${\rm Null}
(\tilde{\textbf{G}}_k)$ is greater than $1$, i.e., $\dim \left( {\rm Null}
(\tilde{\textbf{G}}_k)\right)> 1$, the transmit precoding vector $\textbf{w}_k$
could be further optimized in the sense of maximizing the term
$\left| {{\textbf{h}_{MM_k}}\textbf{w}_k} \right|^2$. This
optimization problem can be formulated as
\begin{equation}\label{opt}
\begin{gathered}
{\mathop{\rm \textbf{w}}\nolimits} _k^{{\mathop{\rm opt}\nolimits} } = \arg \max {\left| {{\textbf{h}_{MM_k}}\textbf{w}_k} \right|^2} \hfill\\
s.t.\quad{\left\| \textbf{w}_k \right\|^2} = 1, \hfill\\
\quad\quad\quad{\textbf{w}_k} \in {\mathop{\rm Null}\nolimits} (\tilde{\textbf{G}}_k).\hfill\\
\end{gathered}
\end{equation}

Denoting $\textbf{C}_k={\rm Null}(\tilde{\textbf{G}}_k)$, the original
optimization problem is equivalent to
\begin{equation}
\begin{gathered}
{{\mathop{\rm \textbf{x}}\nolimits}_{{\rm opt}}} = \arg \max {\left| {{\textbf{h}_{MM_k}}\textbf{C}_k\textbf{x}} \right|^2} \hfill\\
s.t.\quad{\left\| \textbf{x} \right\|^2} = 1,\hfill\\
\end{gathered}
\end{equation}
where $\textbf{x}$ satisfies $\textbf{w}_k=\textbf{C}_k\textbf{x}$.
This problem is convex with ${{\mathop{\rm \textbf{x}}\nolimits}
_{\rm{opt}}} =
\frac{{{{({\textbf{h}_{MM_k}}\textbf{C}_k)}^H}}}{{\left\|
{{\textbf{h}_{MM_k}}\textbf{C}_k} \right\|}}$, i.e.,

\begin{equation}
{\mathop{\rm
\textbf{w}}\nolimits} _k^{{\mathop{\rm opt}\nolimits} } =
\textbf{C}_k\frac{{{{({\textbf{h}_{MM_k}}\textbf{C}_k)}^H}}}{{\left\|
{{\textbf{h}_{MM_k}}\textbf{C}_k} \right\|}}.
\end{equation}

Following~\cite{andrews}, ${\left| {{\textbf{h}_{MM_k}}{\mathop{\rm
\textbf{w}}\nolimits} _k^{{\mathop{\rm opt}\nolimits} }} \right|^2}
\sim \chi _{2({N_B} - (K + M - 1))}^2$, ${\left| {{h_{{R_i}M_k}}}
\right|^2} \sim \chi _{R_{i}M_k}^2(2)$. As a result, the received
SINR in (\ref{start}) under IC is statistically equivalent to
\begin{equation}
\gamma _{_{M{M_k}}}^{IC} \sim \frac{{{P_M}\chi _{2({N_B} -
(K + M - 1))}^2}}{{{P_R}\chi _{2M}^2}}.
\end{equation}

Before starting the performance analysis, we present the following lemmas.

\textbf{\emph{Lemma 1:}} Consider independent random variables (RVs) $X \sim \chi _{2L}^2$ and $Y \sim \chi
_{2M}^2$. The cumulative
distribution function (CDF) of $Z = \frac{X}{aY+b}$ is
\begin{equation}
\begin{gathered}
{F_Z}(z) = 1 - \frac{{{e^{ - bz}}}}{{(M - 1)!}}\sum\limits_{k =
0}^{L - 1} {\frac{{{{(az)}^k}}}{{k!}}} \sum\limits_{i = 0}^k {C_k^i}
{\left(\frac{b}{a}\right)^i} \hfill\\
\quad\quad\quad{(az + 1)^{ - (k + M - i)}}\Gamma (k + M - i).\hfill
\end{gathered}
\end{equation}

\emph{Proof:} See Appendix A.

According to \emph{Lemma 1}, the CDF of ${\gamma _{M{M_k}}^{IC}}$
can be directly derived as
\begin{equation}\label{ICM}
P_{{\gamma _{M{M_k}}}}^{IC}(x)\!\! =\! 1\! - \frac{1}{{(M \!-
\!1)!}}\sum\limits_{k = 0}^{{N_B}\! -\!K \!-\! M}
{\frac{{{{(az)}^k}}}{{k!}}}
 {(az\! \!+\! 1)^{ \!-\!(k\! +\! M)}}\Gamma\!(k\! +\!M),
\end{equation}
where $a = \frac{{{P_R}}}{{{P_M}}}$. Meanwhile, the CDF of
${\gamma _{R{R_i}}^{IC}}$ follows the chi-square distribution, i.e.,
\begin{equation}\label{ICR}
{P^{IC}_{{\gamma _{R{R_i}}}}}(x) = 1- e^{-\frac{x}{P_{R}}}.
\end{equation}

\subsection{Beamforming (BF)}

In the single-cell scenario, eigen-beamforming is optimal for
the multiple-input single-output system~\cite{MISO}. For the $k$-th user, the precoding matrix $\textbf{w}_k$ can be
expressed as $\textbf{w}_k = \frac{{{\textbf{h}_{MM_k}^H}}}{{\left\|
{{\textbf{h}_{MM_k}}} \right\|}}$. Therefore, we can have ${\left|
{{\textbf{h}_{MM_k}}\textbf{w}_k} \right|^2} \sim \chi _{2N_{B}}^2$.
Accounting for the term ${{\left|
{{\textbf{h}_{MM_{k}}}\textbf{w}_j} \right|}^2}$, since the design
of the precoder $\textbf{w}_j$ is independent of
$\textbf{h}_{MM_{k}}$ and $\textbf{w}_j$ is a normalized vector with
unit-norm, we can easily get ${{\left|
{{\textbf{h}_{MM_{k}}}\textbf{w}_j} \right|}^2} \sim \chi_{2}^2$.
The $\gamma _{MM_k}$ in (\ref{start}) is statistically
equivalent to
\begin{equation}\label{BFMM}
\gamma _{_{M{M_k}}}^{BF} \sim \frac{{{P_M}\chi
_{2{N_B}}^2}}{{{P_M}\chi _{2({K-1})}^2 + {P_R}\chi _{2M}^2}}.
\end{equation}

Since $\textbf{g}_{M{R_{i}}}$ and $\textbf{w}_j$ are
independent and $\|\textbf{w}_j\|^2 = 1$, we also have ${\left|
{{\textbf{g}_{M{R_i}}}\textbf{w}_j} \right|^2} \sim \chi
_{MR_{i}}^2(2)$. Accordingly, we have
\begin{equation}
{\gamma _{R{R_i}}^{BF}} \sim \frac{{{P_R} \cdot \chi _{2}^2}}{{{P_M}
\cdot \chi _{2K}^2 + 1}}.
\end{equation}

\textbf{\emph{Lemma 2:}} Consider independent RVs $X \sim \chi _{2L}^2$, $Y_1 \sim \chi _{2M}^2$, and $Y_2 \sim \chi _{2N}^2$. The CDF of $Z = \frac{X}{aY_1+bY_2}$ is
\begin{equation}\label{FZz}
{F_Z}(z) = \int_0^z {\frac{{{a^M}{b^N}{x^{L - 1}}}}{{\Gamma
(L)\Gamma (M + N)}}I(M,N,L,a,b,x) dx},
\end{equation}
where
\begin{equation}
\begin{gathered}
I(M,N,L,a,b,x) = \int_0^\infty {{x^{M + N + L - 1}}{e^{ - (y +
\frac{1}{b})y}}} \hfill\\
\quad\quad\quad\quad{{}_1{F_1}\left(M;N + M; - (\frac{1}{a} -
\frac{1}{b})y\right)} dy.\hfill\\
\end{gathered}
\end{equation}

The expression (\ref{FZz}) can be approximately obtained as
\begin{equation}
\begin{gathered}
{F_Z}(z) \approx 1 - \frac{{{e^{ - bNz}}}}{{(M -
1)!}}\sum\limits_{k = 0}^{L - 1} {\frac{{{{(az)}^k}}}{{k!}}}
\sum\limits_{i = 0}^k {C_k^i} {\left(\frac{bN}{a}\right)^i} \hfill\\
\quad\quad\quad\quad{(az + 1)^{ - (k + M - i)}}\Gamma (k + M - i). \hfill\\
\end{gathered}
\end{equation}

\emph{Proof:} See Appendix B.

Here, by comparing the SINR distribution in (\ref{BFMM}) with the RVs defined in \emph{Lemma 2},
the CDF of ${\gamma _{M{M_k}}^{BF}}$ can be approximately expressed as
\begin{equation}\label{BFM}
\begin{gathered}
P_{{\gamma _{M{M_k}}}}^{BF}(x) \approx 1 - \frac{{{e^{ -
(K-1)x}}}}{{(M - 1)!}}\sum\limits_{k = 0}^{N_B - 1}
{\frac{{{{(ax)}^k}}}{{k!}}} \sum\limits_{i = 0}^k {C_k^i}
{\left(\frac{K-1}{a}\right)^i} \hfill\\
\quad\quad\quad\quad\quad{(ax + 1)^{ - (k + M - i)}}\Gamma (k + M - i),\hfill\\
\end{gathered}
\end{equation}
where $a = \frac{P_R}{P_M}$. Meanwhile, according to
\emph{Lemma 1}, the CDF of ${\gamma _{R{R_i}}^{BF}}$ can be
directly obtained as
\begin{equation}\label{BFR}
P_{{\gamma _{R{R_i}}}}^{BF}(x) = 1 - {e^{ - bx}}{\Big(\frac{x}{a} + 1\Big)^{
- K}},
\end{equation}
where $b = \frac{1}{{{P_R}}}$.

Compared with IC, the above analytical results suggest
that the received signal power at an MUE under BF changes from
a $\chi _{2({N_B} - (M + K - 1))}^2$ RV to a $\chi _{2{N_B}}^2$ RV
with increased degrees of freedom (DoFs). Meanwhile, the RUE
interference power is increased from 0 to a ${\chi_{2K} ^2}$ RV. Thus the effects of the precoding schemes on the system performance are not immediately clear.
In the following sections, three performance metrics, i.e., outage probability, average BER, and system capacity, are characterized when the MBS employs these precoding schemes.

\section{Performance Analysis of Precoding Schemes}

In this section, we analyze system performance gains under the two precoding schemes. From the distribution of the SINR for the MUE and RUE, we see that the interference experienced by the RUE is eliminated at the expense of sacrificing the spatial degrees of freedom of the MBS. Therefore, the effects of the precoding schemes on system performance is the focus of the following paragraphs, i.e., we characterize the outage probability, system sum capacity and average BER when the MBS uses the different precoding schemes.

\subsection{Overall Outage Probability}

A system outage occurs when any received SINR of any potential link for the MBS-association and RRH-association falls below a threshold SINR. We use the overall outage probability to evaluate the performance of these two precoding schemes~\cite{BFIC,BF,andrews}, which can be formulated as
\begin{equation}\label{Pout}
\begin{gathered}
{P_{out}} = \Pr \{ \min ({\gamma _{MM_1}}, \cdots ,{\gamma _{MM_K}}, {\gamma _{R{R_1}}}, \cdots ,{\gamma _{R{R_M}}})\textless {\gamma _{th}}\} \hfill\\
\quad\quad= 1 - \Pr \{ {\gamma _{MM_1}} \textgreater {\gamma _{th}}, \cdots , {\gamma _{MM_K}} \textgreater {\gamma _{th}}, \hfill\\
\quad\quad\quad{\gamma _{R{R_1}}} \textgreater {\gamma _{th}}, \ldots ,{\gamma _{R{R_M}}} \textgreater {\gamma _{th}}\}, \hfill\\
\end{gathered}
\end{equation}
where ${\gamma _{th}}$ is the SINR threshold.

Considering that all elements of the channels for the various pairs of
transmitters and receivers are independent, (\ref{Pout}) can be
rewritten as
\begin{equation}\label{Pout1}
\begin{gathered}
{P_{out}} = 1 - \prod\limits_{k = 1}^K {\Pr \{ {\gamma _{MM_k}}{\rm{
> }}{\gamma _{th}}\} } \prod\limits_{i = 1}^M {\Pr \{ {\gamma
_{R{R_i}}}{\rm{ > }}{\gamma _{th}}\} } \hfill\\
\quad\quad = 1 - [1 - {P_{{\gamma
_{MM_k}}}(\gamma _{th})}]^K[1 - {P_{{\gamma _{RR_i}}}(\gamma
_{th})}]^M. \hfill\\
\end{gathered}
\end{equation}

Due to the substantial differences between the aforementioned
$P_{{\gamma _{MM_k}}}$ and $P_{{\gamma _{RR_i}}}$, closed-form
expressions for $P_{out}$ with the two precoding schemes are presented
independently as follows.

\textbf {IC}: Substituting (\ref {ICM}) and (\ref {ICR}) into (\ref
{Pout1}), the overall outage probability of the IC scheme in H-CRANs
can be derived as
\begin{align}\label{PBF1}
P_{out}^{IC}\!=&\!1\!-\!\bigg[\frac{1}{(M\!-\!1)!}\sum_{k\!=\!0}^{N_{B}
-K-M}\frac{(a\gamma_{th})^{k}}{k!}(a\gamma_{th}+1)^{-(k+M)}\nonumber\\
&\Gamma(k\!+\!M)\bigg]^{K}e^{-\frac{M\gamma_{th}}{P_{R}}},
\end{align}
where $a=\frac{P_R}{P_M}$.

\textbf {BF}: Substituting (\ref {BFM}) and (\ref {BFR}) into (\ref
{Pout1}), the overall outage probability of the BF scheme in H-CRANs
can be derived as
\begin{align}\label{PIC1}
P_{out}^{BF} & =\!1\!-\!\bigg[\frac{e^{-(K\!-\!1)\gamma_{th}}}{(M\!-\!1)!}\sum_{k\!=\!0}^{N_{B}
-1}\frac{(a\gamma_{th})^{k}}{k!}\sum_{i\!=\!0}^{k}C_{k}^{i}(\frac{K\!-\!1}{a})^i\nonumber\\
&\!\!\!\!(a\gamma_{th}\!+\!1)^{-\!(k\!+M\!-\!i)}\Gamma(k\!+\!M\!\!-\!i)\bigg]^{K}\bigg[e^{-\frac{x}{P_{R}}}(\frac{\gamma_{th}}{a}\!+\!1)\bigg]^{-KM}.
\end{align}

Both (\ref {PBF1}) and (\ref {PIC1}) show that the overall outage probability strictly depends on $N_B$, $K$, $M$, $\gamma_{th}$, and the ratio of $P_R$ to $P_M$. It is hard to directly determine which precoder is better, and thus we will show performance comparisons for these two methods when taking different configurations into account. The precoding scheme is adaptively selected to minimize the overall outage probability.

\subsection{Sum Capacity}

The sum capacity of the entire system can be expressed as
\begin{equation}\label{R}
R = \sum\limits_{k = 1}^K {\mathbb{E}\big[\log_2(1 + {\gamma _{M{M_k}}})\big]} + \sum\limits_{i = 1}^M {\mathbb{E}\big[\log_2(1 + {\gamma _{R{R_i}}})\big]}.
\end{equation}

Before analyzing the sum capacity under IC and BF schemes, we present the following
lemmas.

\textbf{\emph{Lemma 3:}} Consider independent RVs $X \sim \chi _{2L}^2$ and $Y \sim \chi _{2M}^2$, and define $Z = \frac{X}{aY + b}$. We have

\begin{equation}\label{R1}
\begin{gathered}
\quad\quad {R_1}(a,b,L,M) \buildrel \Delta \over = {\mathbb{E}}[{\log_2}(1 + Z)] \hfill\\
= \frac{1}{{{\mathop{\ln}\nolimits} 2(M - 1)!}}\sum\limits_{k = 0}^{L - 1} {\frac{{{{(a)}^{i - M}}}}{{k!}}} \sum\limits_{i = 0}^k {C_k^i} {(\frac{b}{a})^i}\Gamma (k + M- i ) \hfill\\
\quad\big[{(\frac{1}{a}\! - \!1)^{i -\! M -\! k}}{e^b}\Gamma (k \!+ \!1)\Gamma ( \!- k,b) \!-\!\! \sum\limits_{j\! =\! 1}^{k\! - \!i + M} {\sum\limits_{m = 0}^k {C_k^m} } {(\! - \frac{1}{a})^m} \hfill\\
\quad\quad {a^{\! -\! k\! + \!m\! +\! j\! -\! 1}}{e^{\frac{b}{a}}}\Gamma (k - i - j + 1,\frac{b}{a}){\kern 1pt} {\kern 1pt} {(\frac{1}{a} - 1)^{i + j- M - k - 1}}\big]. \hfill\\
\end{gathered}
\end{equation}

\emph{Proof:}
Given two independent RVs $X \sim \chi _{2L}^2$, $Y \sim \chi _{2M}^2$, and $a \textgreater 0, b \textgreater 0$, by defining $Z = \frac{X}{aY + b}$, its CDF can be expressed as~\cite{BER}

\begin{equation}\label{1}
\begin{gathered}
{F_Z}(z) =  \int_0^\infty  {{F_X}(ayz + bz)} {f_Y}(y)dy \hfill\\
\quad\quad = 1 - \frac{{{e^{ - bz}}}}{{(M -
1)!}}\sum\limits_{k\! = \!0}^{L - 1}
\sum\limits_{i = 0}^k {C_k^i} {\frac{{{{(az)}^k}}}{{k!}}}{(\frac{b}{a})^i} \hfill\\
\quad\quad\quad\quad  {(az + 1)^{- (k + M - i)}}\Gamma (k + M - i).\hfill\\
\end{gathered}
\end{equation}

With this CDF expression, we have

\begin{equation}\label{Ez}
\begin{gathered}
{\mathbb{E}}[{\log_2}(1 + z)]{\kern 1pt} = \frac{1}{{{\mathop{\ln}} 2}}\int_0^\infty  {\frac{{1 - {F_Z}(z)}}{{1 + z}}} dz \hfill\\
\quad\quad\quad\quad=  \frac{1}{{{\mathop{\ln}\nolimits} 2(M - 1)!}}\sum\limits_{k = 0}^{L - 1} {\frac{{{{(a)}^{i - M}}}}{{k!}}} \sum\limits_{i = 0}^k {C_k^i} {(\frac{b}{a})^i} \hfill\\
 \quad\quad\quad\quad  \Gamma (k + M - i)\int_0^\infty  {\frac{{{z^k}{e^{ - bz}}}}{{(z + 1){{(z + \frac{1}{a})}^{k + M - i }}}}}dz.\hfill\\
\end{gathered}
\end{equation}

By applying the decomposition

\begin{equation}\label{de}
\begin{gathered}
\quad \frac{1}{{(z + 1){{(z + \frac{1}{a})}^{k + M - i}}}} \hfill\\
 = \frac{{{{(\frac{1}{a} - 1)}^{i - M - k}}}}{{z + 1}} - \sum\limits_{j = 1}^{k - i + M} {\frac{{{{(\frac{1}{a} - 1)}^{i + j - M - k - 1}}}}{{{{(z + \frac{1}{a})}^j}}}}, \hfill\\
\end{gathered}
\end{equation}
the ergodic capacity in (\ref{Ez}) can be rewritten as

\begin{equation}\label{add2}
\begin{gathered}
\quad\quad {\mathbb{E}}[\log_2(1 + z)]\hfill\\
= \frac{1}{{{\mathop{\rm In}\nolimits} 2(M - 1)!}}\sum\limits_{k = 0}^{L - 1} {\frac{{{{a}^{i - M}}}}{{k!}}} \sum\limits_{i = 0}^k {C_k^i} {(\frac{b}{a})^i}\Gamma (k  + M - i) \hfill\\
\quad \big[\int_0^\infty  {\frac{{{{(\frac{1}{a} - 1)}^{i - M - k}}{z^k}{e^{ - bz}}}}{{z + 1}}} dz - \sum\limits_{j = 1}^{k - i + M} {\sum\limits_{m = 0}^k {C_k^m} } \hfill\\
\quad {{(\frac{1}{a} - 1)}^{i + j - M - k - 1}} {( - \frac{1}{a})^m}\int_0^\infty  {{{(z + \frac{1}{a})}^{k - m - j}}{e^{ - bz}}} dz\big] \hfill\\
\end{gathered}
\end{equation}
\begin{equation}\label{add3}
\begin{gathered}
= \frac{1}{{{\mathop{\rm In}\nolimits} 2(M - 1)!}}\sum\limits_{k = 0}^{L - 1} {\frac{{{{a}^{i - M}}}}{{k!}}} \sum\limits_{i = 0}^k {C_k^i} {(\frac{b}{a})^i}\Gamma (k  + M - i) \hfill\\
\quad\big[{(\frac{1}{a} - \!1)^{i -\! M -\! k}}{e^b}\Gamma (k \!+ 1)\Gamma ( \!- k,b) - \sum\limits_{j\! = 1}^{k\! - \!i + M} {\sum\limits_{m = 0}^k {C_k^m} } {( - \!\frac{1}{a})^m} \hfill\\
\quad\quad {a^{ - k + m + j - 1}}{e^{\frac{b}{a}}}\Gamma (k - i - j + 1,\frac{b}{a}){\kern 1pt} {\kern 1pt} {(\frac{1}{a} - 1)^{i + j - M - k - 1}}\big], \hfill\\
\end{gathered}
\end{equation}
where (\ref{add2}) is obtained by performing binomial expansion on the term $(z+\frac{1}{a}-\frac{1}{a})^k$. Then (\ref{add3}) is
obtained according to Eq. 3.383.10 and Eq. 3.382.4 in~\cite{Table}.

\textbf{\emph{Lemma 4:}} For an RV $X \sim \chi _{2}^2$, and $Y = \delta X, \delta>0$, we have
\begin{equation}\label{R2}
{R_2}(\delta ) \buildrel \Delta \over = {\mathbb{E}}[{\log_2}(1 + Y)] = \frac{1}{{{\ln} 2}}{e^{\frac{1}{\delta }}}{\mathbb{E}_1}\bigg(\frac{1}{\delta }\bigg),
\end{equation}
where ${\mathbb{E}_1}(z) = \int_z^\infty {\frac{{{e^{- t}}}}{t}} dt$ is the exponential integral function of the first order.

\emph{Proof:} According to the chi-squared distribution, we have the CDF of $Y$ is
\begin{equation}
{F_Y}(y) = 1 - {e^{ - \frac{y}{\delta }}},
\end{equation}

and thus
\begin{equation}
\begin{gathered}
\mathbb{E}[\log_2(1 + Y)] = \frac{1}{{{\mathop{\ln}\nolimits} 2}}\int_0^\infty {\frac{{1 - {F_Y}(y)}}{{1 + y}}} dy = \hfill\\ \quad\quad\quad \frac{1}{{{\mathop{\ln}\nolimits} 2}}\int_0^\infty {\frac{{{e^{ - \frac{y}{\delta }}}}}{{1 + y}}} dy = \frac{1}{{{\mathop{\ln}\nolimits} 2}}{e^{\frac{1}{\delta }}}{\mathbb{E}_1}\bigg(\frac{1}{\delta }\bigg).
\end{gathered}
\end{equation}

Due to the differences between the aforementioned ${\gamma _{MM_k}}$
and ${\gamma _{RR_i}}$, using Lemma 3 and Lemma 4, closed-form expressions for sum capacity $R$
under these two precoding schemes are presented as follows.
\begin{itemize}
 \item \textbf{IC:} Substituting the distribution of $\gamma _{M{M_k}}^{IC}$ and $\gamma _{R{R_i}}^{IC}$ into (\ref{R}), the sum capacity under the IC scheme in H-CRANs can be derived as
 \begin{equation}\label{RIC}
 \begin{split}
 R^{IC} ={R_1}\Big(\frac{P_R}{P_M},0,N_B-M,M\Big)+{R_2}\Big(P_R\Big), \hfill\\
 \end{split}
 \end{equation}
 where ${R_1}(\cdot)$ and ${R_2}(\cdot)$ follow (\ref{R1}) and (\ref{R2}).
 \item \textbf{BF:} Substituting the distribution of $\gamma _{M{M_k}}^{BF}$ and $\gamma _{R{R_i}}^{BF}$ into (\ref{R}), the
 sum capacity under the BF scheme in H-CRANs can be derived as
 \begin{equation}\label{PBF}
 \begin{split}
 R^{BF}={R_1}\Big(\frac{P_R}{P_M},\!K-1,N_B,M\Big)+{R_1}\Big(\frac{P_M}{P_R},\frac{1}{P_R},1,K\Big), \hfill\\
 \end{split}
 \end{equation}
 where ${R_1}(\cdot)$ follows (\ref{R1}).
\end{itemize}

Similar to the overall outage probability results in (\ref {PBF1}) and (\ref {PIC1}), the derived sum capacity strictly depends on $N_B$, $K$, $M$, and the ratio of $P_R$ to $P_M$. It is hard to directly judge which precoder is better, as this depends on the specific system configurations.

\subsection{Average Bit Error Rate}

The average BER is defined as the average BER of all radio links, which can be expressed as
\begin{equation}
{B_e} \buildrel \Delta \over = \frac{1}{K + M}\bigg(\sum\limits_{k = 1}^K {B_M^k + \sum\limits_{i = 1}^M {B_R^i} }\bigg ),
\end{equation}
where $B_M^k$ is the BER of the link between the MBS and the $k$-th MUE, and $B_R^i$ is the BER of the RRH-RUE link in the $i$-th cell.

Note that the average BER of two
end nodes is dominated by the worst one~\cite{BER}; therefore, we can rewrite the
average BER approximately as
\begin{equation}
{B_e} \approx \frac{1}{K + M}\max \{B_M^1, \cdots, B_M^K, B_R^1, \cdots, B_R^M\}.
\end{equation}

For commonly used modulation schemes, the BER of each link $B_b$ can be written in the form
\begin{equation}
{B_b} = \mathbb{E}[{\beta_1}Q(\sqrt {2{\beta_2}\gamma})] = \int_0^{ +
\infty }{{\beta _1}Q(\sqrt{2{\beta_2}z} }) {p_\gamma }(z)dz,
\end{equation}
where ${\beta _1}$ and ${\beta _2}$ are coefficients that depend on the modulation mode, and $Q(x) = \frac{1}{{\sqrt {2\pi } }}\int_x^\infty {\exp(-
\frac{u^2}{2})}du$
is the tail probability of the standard normal distribution.

For simplicity, we consider Binary Phase Shift Keying (BPSK) modulation in this paper, which
corresponds to ${\beta_1} = {\beta_2} = 1$. For other modulation formats,
similar results could also be obtained. Note that $Q(x)$ is monotonically
decreasing when $x \ge 0$. Thus, the average BER can be approximated as
\begin{equation}\label{Pe}
{B_e} \approx \frac{1}{{2(K + M)\sqrt \pi }}\int_0^{ + \infty } {\frac{{{e^{ - z}}}}{{\sqrt z }}} {P_{{\gamma _e}}}(z)dz,
\end{equation}
where
\begin{equation}
{\gamma _e} = \min \{ {\gamma _{M{M_1}}}, ..., {\gamma _{M{M_K}}}, {\gamma _{R{R_1}}}, ..., {\gamma _{R{R_M}}}\}. \hfill\\
\end{equation}

The CDF of ${\gamma _e}$ can be expressed as
\begin{equation}\label{PoutCDF}
\begin{split}
{P_{{\gamma _e}}}(z) &= \Pr \{\min({\gamma _{MM_1}},...,{\gamma _{MM_K}}, {\gamma _{R{R_1}}},...,{\gamma _{R{R_M}}})\! <\! z\} \hfill \\
&= 1 - \big[1 - {P_{{\gamma _{MM_k}}}(z)}\big]^K\big[1 - {P_{{\gamma _{RR_i}}}(z)}\big]^M.\
\end{split}
\end{equation}

Due to aforementioned differences in ${\gamma_{MM_k}}$ and ${\gamma_{RR_i}}$, the expressions for $B_e$ under the two schemes
are presented separately as follows.
\begin{itemize}
 \item \textbf{IC:} Substituting (\ref {ICM}) and (\ref {ICR}) into (\ref{PoutCDF}), and further into (\ref{Pe}), the average BER under the IC scheme can be obtained as
 \begin{equation}\label{IC_BER}
 {B^{IC}_e} \approx \frac{1}{{2(K + M)\sqrt \pi }}\int_0^{ + \infty } {\frac{{{e^{ - z}}}}{{\sqrt z }}} {P_{{\gamma _e}}^{IC}}(z)dz,
 \end{equation}
 where
 \begin{align}\label{PBF}
 P_{\gamma _e}^{IC}(z)= &1\!-\!\bigg[\frac{1}{(M\!-\!1)!}\sum_{k=0}^{N
 -K-M}\frac{(az)^{k}}{k!}(az+1)^{-(k+M)}\nonumber\\
 &\Gamma(k+M)\bigg]^{K}e^{-\frac{Mz}{P_{R}}}
 \end{align}
 with $a=\frac{P_R}{P_M}$.

 \item \textbf{BF:} Substituting (\ref {BFM}) and (\ref {BFR}) into (\ref{PoutCDF}), and further into (\ref{Pe}), the average BER under the BF scheme can be obtained as
 \begin{equation}\label{BF_BER}
 {B^{BF}_e} \approx \frac{1}{{2(K + M)\sqrt \pi }}\int_0^{ + \infty } {\frac{{{e^{ - z}}}}{{\sqrt z }}} {P_{{\gamma _e}}^{BF}}(z)dz,
 \end{equation}
 where
 \begin{align}\label{PIC}
 P_{\gamma _e}^{BF}(z) =\!1\!-\!\bigg[\frac{e^{-(K\!-\!1)z}}{(M\!-\!1)!}\sum_{k\!=\!0}^{N_{B}
 \!-\!1}\frac{(az)^{k}}{k!}\sum_{i\!=\!0}^{k}C_{k}^{i}(\frac{K\!-\!1}{a})^i\quad\nonumber\\
 (az\!+\!1)^{-(k\!+M\!-\!i)}\Gamma(k\!+\!M\!-\!i)\bigg]^{K}\bigg[e^{-\frac{x}{P_{R}}}(\frac{z}{a}\!+\!1)\bigg]^{-KM}.
 \end{align}
\end{itemize}

The aforementioned expressions for IC and BF suggest that the average BER depends on the system configuration, such as the number of RRHs \emph{M}, the number of MUEs \emph{K}, the transmit power per antenna in the MBS $P_M$, and the transmit power per antenna in the RRH $P_R$. It is hard to directly compare which is better between (\ref {IC_BER}) and (\ref {BF_BER}), and such comparisons will be based on the numerical results shown in Section VI.

\section{Inter-tier Cooperative Radio Resource Allocation}

Under the inter-tier collaborative precoding schemes IC and BF, CRRA can be used to further suppress the inter-tier interference to optimize the throughput of H-CRANs.
Since the MUE prefers to access the MBS for seamless coverage, while the RUE often associates with RRHs to achieve high bit rate, we can maximize the RUEs' aggregated rates while guaranteeing the MUEs' summarized bit rates to model the throughput maximization problem. Furthermore, we can assume that the power of each RRH is different to make the power allocation for each RRH flexible. Hence, the throughput maximization problem for H-CRANs can be formulated as
\begin{equation}\label{opt0}
\begin{split}
\max \quad & \mathcal{R}_R = {\sum\limits_{i = 1}^M {\log_2(1 + {\gamma _{R{R_i}}})}}\\
s.t.\quad & P_M\leq P_{MS},\hfill\\
& P_{R_i}\leq P_{RS_i}, \quad i = 1,2,...,M, \hfill\\
&\sum \limits_{i=1}^{M} P_{R_i}\leq P_{RS}, \hfill\\
&{\log_2(1 + {\gamma _{M{M_k}}})} \geq \mathcal{R}_{MS}, \quad k=1,2,...,K,\hfill\\
\end{split}
\end{equation}
where $\mathcal{R}_{MS}$ is the QoS threshold of each MUE, $P_{MS}$ and $P_{RS_i}$ are respectively
the power limits of the MBS and RRH $i$, and $P_{RS}$ is the total power threshold of the RRHs.
Note that problem (\ref{opt0}) is feasible
only if the following condition holds for the $k$-th MUE:
\begin{equation}\label{bitrate_thre}
2^{\mathcal{R}_{MS}}-1 \leq \frac{P_{MS} \gamma _{M{M_k}}}{P_M}.
\end{equation}

This condition indicates that the QoS thresholds for
the MUEs should not be too high. This can be intuitively understood since the QoS constraints must at least
be satisfied when the maximum allowable power of the MBS $P_{MS}$ is applied. Therefore,
throughout the rest of the paper, we assume that (\ref{bitrate_thre})
always holds for any MUE, which guarantees that an optimal power allocation exists.

\subsection{Interference Collaboration}

When the IC scheme is used at the MBS, the precoding vector is chosen to eliminate
the inter-tier interference to other MUEs and RUEs, i.e.,
\begin{equation}\label{w_opt}
{\mathop{\rm
\textbf{w}}\nolimits} _k^{{\mathop{\rm opt}\nolimits} } =
\textbf{C}_k\frac{{{{({\textbf{h}_{MM_k}}\textbf{C}_k)}^H}}}{{\left\|
{{\textbf{h}_{MM_k}}\textbf{C}_k} \right\|}}.
\end{equation}

Hence, $\gamma _{RR_i}$ in (\ref{RR}) can be substituted into (\ref{opt0}), and the transmit bit rate threshold for the MUE $\mathcal{R}_{MS}$ can be further derived if (\ref{w_opt}) is substituted into $\gamma
_{MM_k}$ in (\ref{RR}). Accordingly, the throughput maximization
problem for the IC based CRRA in (\ref{opt0}) can be reformulated as
\begin{equation}\label{optIC}
\begin{split}
\mathop {\max }\limits_{ \{P_M, P_{R_i}\} } \quad & \mathcal{R}_R^{IC} ={\sum\limits_{i = 1}^M {\log_2(1 + {P_{R_i}}{\left| {g_{R{R_i}}} \right|^2})}} \hfill\\
s.t.\quad & P_M\leq P_{MS},\hfill\\
& P_{R_i}\leq P_{RS_i}, \quad i = 1,2,...,M, \hfill\\
&\sum \limits_{i=1}^{M} P_{R_i}\leq P_{RS}, \hfill\\
&{\log_2(1 + {\gamma _{M{M_k}}})} \geq \mathcal{R}_{MS}, \quad k=1,2,...,K,\hfill\\
&{\gamma _{MM_k}^{IC}} = \frac{{{P_M}{{\left|
{{\textbf{h}_{MM_k}}\textbf{w}_k^{{\mathop{\rm opt}\nolimits} }} \right|}^2}}}{{ \sum_{i = 1}^M {{P_{R_i}}{{\left|{{h_{{R_i}M_k}}} \right|}^2}}}}, \quad k=1,2,...,K. \hfill\\
\end{split}
\end{equation}

Noting that the RUEs' sum rates $\mathcal{R}_R^{IC}$ are only determined by the RRH's power $P_{R_i}$,
we have the following proposition.

\textbf{\emph{Proposition 1:}}
Let $\{P_M^{{\mathop{\rm opt}\nolimits} }, P_{R_i}^{{\mathop{\rm opt}\nolimits} }\}$ denote the solution to problem (\ref{optIC}).
Define $A_k=\frac{{{P_{MS}}{{\left|
{{\textbf{h}_{MM_k}}\textbf{w}_k^{{\mathop{\rm opt}\nolimits} }} \right|}^2}}}{2^{\mathcal{R}_{MS}}-1}$.
Then $P_M^{{\mathop{\rm opt}\nolimits} } = P_{MS}$, and
\begin{equation}
P_{R_i}^{{\mathop{\rm opt}\nolimits} }= \frac{1}{\lambda_i + \mu + \sum_{k = 1}^K \nu_k{{\left|{{h_{{R_i}M_k}}} \right|}^2}}-\frac{1}{{\left| {g_{R{R_i}}} \right|^2}},
\end{equation}
where $\lambda_i,\mu,\nu_k$ are chosen elaborately such that (49)-(54).

\emph{Proof:} Since
the MBS power $P_M$ occurs only in the constraints and does not affect the RUEs' sum rates $\mathcal{R}_R^{IC}$ in the IC scheme,
an optimal $P_M$ can be achieved in the following limitation under a fixed $P_{R_i}$:
\begin{equation}
\Big[\frac{(2^{\mathcal{R}_{MS}}-1)\sum\nolimits_{i = 1}^M P_{R_i} {\left|{h_{R_i} M_k} \right|}^2}{{\left| {{\textbf{h}_{MM_k}}\textbf{w}_k^{{\mathop{\rm opt}\nolimits} }} \right|}^2},P_{MS}\Big].
\end{equation}

However, considering the maximization of the MBS's coverage, we denote the optimal $P_M$
as $P_M^{{\mathop{\rm opt}\nolimits} }=P_{MS}$. Substituting $P_M^{{\mathop{\rm opt}\nolimits}}$ into the original problem (\ref{optIC}), (\ref{optIC})
can be simplified to the following problem:
\begin{equation}\label{optIC1}
\begin{split}
\mathop {\max }\limits_{ \{P_{R_i}\} } \quad & {\sum\limits_{i = 1}^M {\log_2(1 + {P_{R_i}}{\left| {g_{R{R_i}}} \right|^2})}} \hfill\\
s.t.\quad &P_{R_i}\leq P_{RS_i}, \quad i = 1,2,...,M, \hfill\\
&\sum \limits_{i=1}^{M} P_{R_i}\leq P_{RS}, \hfill\\
&{{ \sum_{i = 1}^M {{P_{R_i}}{{\left|{{h_{{R_i}M_k}}} \right|}^2}}}} \leq A_k, \quad k=1,2,...,K.\hfill\\
\end{split}
\end{equation}

It is observed that (\ref{optIC1}) is a typical convex optimization problem, which can be solved by employing KKT conditions.
Therefore, the Lagrangian function of (\ref{optIC1}) is defined as
\begin{equation}
\begin{split}
L(P_{R_i},\lambda_i,\mu,\nu_k)=&\ln 2 {\sum\limits_{i = 1}^M {\log_2(1 + {P_{R_i}}{\left| {g_{R{R_i}}} \right|^2})}} \hfill\\
&+\sum_{i = 1}^M \lambda_i (P_{RS_i}-P_{R_i})+\mu (P_{RS}-\sum \limits_{i=1}^{M} P_{R_i}) \hfill\\
&+\sum_{k = 1}^K \nu_k(A_k-{{ \sum_{i = 1}^M {{P_{R_i}}{{\left|{{h_{{R_i}M_k}}} \right|}^2}}}}), \hfill\nonumber\\
\end{split}
\end{equation}
where $\lambda_i \geq 0$, $\mu \geq 0$, and $\nu_k \geq 0$ are the non-negative Lagrange multipliers associated
with the constraints in (\ref{optIC1}). The KKT conditions can be applied on the Lagrangian function to obtain
\begin{align}\label{ICsol}
\frac{\partial L(P_{R_i},\lambda_i,\mu,\nu_k)}{\partial P_{R_i}}=& \frac {{\left| {g_{R{R_i}}} \right|^2}}{1 + {P_{R_i}^{\mathop{\rm opt}\nolimits}}{\left| {g_{R{R_i}}} \right|^2}}
- \lambda_i^{\mathop{\rm opt}\nolimits} \hfill\nonumber\\
 &- \mu^{\mathop{\rm opt}\nolimits} - \sum_{k = 1}^K \nu_k^{\mathop{\rm opt}\nolimits}{{\left|{{h_{{R_i}M_k}}} \right|}^2}\hfill\nonumber\\
= &\quad0, \hfill
\end{align}
\begin{align}
&\lambda_i^{\mathop{\rm opt}\nolimits} (P_{RS_i}-P_{R_i}^{\mathop{\rm opt}\nolimits})=0, \quad i = 1,2,...,M,\hfill\\
&P_{R_i}^{\mathop{\rm opt}\nolimits} \leq P_{RS_i}, \quad i = 1,2,...,M, \hfill\\
&\mu^{\mathop{\rm opt}\nolimits} (P_{RS}-\sum \limits_{i=1}^{M} P_{R_i}^{\mathop{\rm opt}\nolimits})=0, \hfill\\
&\sum \limits_{i=1}^{M} P_{R_i}^{\mathop{\rm opt}\nolimits}\leq P_{RS}, \hfill\\
&\nu_k^{\mathop{\rm opt}\nolimits}(A_k-{{ \sum_{i = 1}^M {{P_{R_i}^{\mathop{\rm opt}\nolimits}}{{\left|{{h_{{R_i}M_k}}} \right|}^2}}}})=0, \quad k=1,2,...,K, \hfill\\
&{{ \sum_{i = 1}^M {{P_{R_i}^{\mathop{\rm opt}\nolimits}}{{\left|{{h_{{R_i}M_k}}} \right|}^2}}}} \leq A_k ,\quad k=1,2,...,K,\hfill
\end{align}
where $\lambda_i^{\mathop{\rm opt}\nolimits}$, $\mu^{\mathop{\rm opt}\nolimits}$, and $\nu_k^{\mathop{\rm opt}\nolimits}$
are optimal solutions to the Lagrangian function.

Based on (\ref{ICsol}), an optimal solution to (\ref{optIC1}) can be obtained, i.e.,

\begin{equation}
P_{R_i}^{\mathop{\rm opt}\nolimits} = \frac{1}{\lambda_i^{\mathop{\rm opt}\nolimits} + \mu^{\mathop{\rm opt}\nolimits} + \sum_{k = 1}^K \nu_k^{\mathop{\rm opt}\nolimits}{{\left|{{h_{{R_i}M_k}}} \right|}^2}}-\frac{1}{{\left| {g_{R{R_i}}} \right|^2}}.
\end{equation}

Note that an optimal solution needs to satisfy (49)-(54).
However, optimal $\lambda_i^{\mathop{\rm opt}\nolimits}$, $\mu^{\mathop{\rm opt}\nolimits}$, and $\nu_k^{\mathop{\rm opt}\nolimits}$ are not easy to find.
Fortunately, $P_{R_i}^{\mathop{\rm opt}\nolimits}$ is monotonically decreasing in each multiplier,
which makes it possible to compute the optimal $\lambda_i^{\mathop{\rm opt}\nolimits}$, $\mu^{\mathop{\rm opt}\nolimits}$, and $\nu_k^{\mathop{\rm opt}\nolimits}$.
The following lemma provides intervals containing the optimal multipliers.

\textbf{\emph{Lemma 5:}} The optimal $\lambda_i^{\mathop{\rm opt}\nolimits}$, $\mu^{\mathop{\rm opt}\nolimits}$, and $\nu_k^{\mathop{\rm opt}\nolimits}$
satisfying (44)-(49) are respectively within $[0,\lambda_i^\textrm{max}]$,
$[0,\mu^\textrm{max}]$, and $[0,\nu_k^\textrm{max}]$,
where $\lambda_i^\textrm{max}={\left| {g_{R{R_i}}} \right|^2}$, $\mu^\textrm{max}=\min_i \{{\left| {g_{R{R_i}}} \right|^2}\}$,
and $\nu_k^\textrm{max}=\min_i \Big\{\frac{\left| {g_{R{R_i}}} \right|^2}{{{\left|{{h_{{R_i}M_k}}} \right|}^2}}\Big\}$.

\emph{Proof:} The results follow from the fact that $P_{R_i}^{\mathop{\rm opt}\nolimits}\geq 0$.

\subsection{Beamforming}

When the BF scheme is used at the MBS, the precoding vector of MUE $k$ is determined by
the $\textbf{h}_{MM_k}$, i.e.,
$\textbf{w}_k = \frac{{{\textbf{h}_{MM_k}^H}}}{{\left\|
{{\textbf{h}_{MM_k}}} \right\|}}$. With precoding vectors $\textbf{w}_k$
fixed, we can obtain the corresponding SINRs for RUE $i$
and MUE $k$, as follows:
\begin{equation}
\begin{split}
{\gamma^{BF} _{R{R_i}}} &= \frac{{{P_{R_i}}{{\left| {{g_{R{R_i}}}} \right|}^2}}}{{{P_M}\sum\limits_{j=1}^K{{\left|
\textbf{g}_{M{R_i}} \textbf{w}_j \right|}^2} + 1}}, \quad i=1,2,...,M, \hfill\\
{\gamma^{BF} _{MM_k}} &= \frac{{{P_M}{{\left| \textbf{h}_{MM_k} \textbf{w}_k \right|}^2}}}{{P_M}\!\sum \limits_{j =
1,j \ne k}^K {{{\left| \textbf{h}_{MM_{k}} \textbf{w}_j
\right|}^2}}\! +\! \sum\limits_{i = 1}^M {P_{R_i}}{{{\left|
{{h_{{R_i}M_{k}}}} \right|}^2}}}, \hfill\\
\qquad k&=1,2,...,K.\hfill\nonumber\\
\end{split}
\end{equation}

Hence the sum rates
optimization problem for the BF scheme can be formulated as
\begin{equation}\label{opt2}
\begin{split}
\mathop {\max }\limits_{ \{P_M, P_{R_i}\} } & \mathcal{R}_R^{BF} ={\sum\limits_{i = 1}^M {\log_2(1 + {\gamma^{BF}_{R{R_i}}})}} \hfill\\
s.t.\quad & P_M\leq P_{MS},\hfill\\
& P_{R_i}\leq P_{RS_i}, \quad i = 1,2,...,M, \hfill\\
&\sum_{i=1}^{M} P_{R_i}\leq P_{RS}, \hfill\\
&{\log_2(1 + {\gamma _{M{M_k}}})} \geq \mathcal{R}_{MS}, \quad k=1,2,...,K.\hfill\\
\end{split}
\end{equation}

Finding the optimal power allocation for such a nonconvex problem is a very challenging
task, since (\ref{opt2}) is not jointly
convex in $\{P_M, P_{R_i}\}$. However, despite this difficulty, we
can provide a stationary solution for (\ref{opt2}) since it is convex in each variable
and can be transformed into a convex problem:

\begin{itemize}
\item[\emph{$\bullet$}] \textbf{Optimal $P_{R_i}$ under fixed $P_M$}:
For fixed $P_M$, $\mathcal{R}_R^{BF}$
is concave in $P_{R_i}$ since
\begin{equation}
\begin{split}
\frac{\partial^2 \{\mathcal{R}_R^{BF}\}}{\partial P_{R_i}^2} = -\frac{C_i^2}{(1+C_i P_{R_i})^2} < 0,\nonumber\hfill
\end{split}
\end{equation}
and (\ref{opt2}) can be simplified into
\begin{equation}\label{optBF1}
\begin{split}
\mathop {\max }\limits_{ \{P_{R_i}\} } \quad & {\sum\limits_{i = 1}^M {\log_2(1 + C_i{P_{R_i}})}} \hfill\\
s.t.\quad & P_{R_i}\leq P_{RS_i}, \quad i = 1,2,...,M, \hfill\\
&\sum_{i=1}^{M} P_{R_i}\leq P_{RS}, \hfill\\
&\sum\limits_{i = 1}^M {P_{R_i}}{{{\left|{{h_{{R_i}M_{k}}}} \right|}^2}} \leq B_k, \quad k=1,2,...,K,\hfill\\
\end{split}
\end{equation}
where
\begin{equation}
\begin{split}
C_i &= \frac{{{\left| {{g_{R{R_i}}}} \right|}^2}}{{{P_M}\sum\limits_{j=1}^K{{\left|
\textbf{g}_{M{R_i}} \textbf{w}_j \right|}^2} + 1}},\quad i = 1,2,...,M,\hfill\\
B_k &=\frac{{P_M}{{\left| \textbf{h}_{MM_k} \textbf{w}_k \right|}^2}}{2^{\mathcal{R}_{MS}}-1}
-{P_M}\!\sum \limits_{j = 1,j \ne k}^K {{{\left| \textbf{h}_{MM_{k}} \textbf{w}_j \right|}^2}},\hfill\\
\quad k& =1,2,...,K.\hfill\nonumber\\
\end{split}
\end{equation}

It is not difficult to see that (\ref{optBF1}) has a similar form to (\ref{optIC1}),
and according to the solution to (\ref{optIC1}), we give the optimal solution to (\ref{optBF1}) as follows:
\begin{equation}
\begin{split}\label{PRIBF}
P_{R_i}^{\mathop{\rm opt}\nolimits} &= \frac{1}{\lambda_i^{\mathop{\rm opt}\nolimits} + \mu^{\mathop{\rm opt}\nolimits} + \sum_{k = 1}^K \nu_k^{\mathop{\rm opt}\nolimits} {{\left|{{h_{{R_i}M_k}}} \right|}^2}}-\frac{1}{C_i},\hfill\\
\end{split}
\end{equation}
with optimal Lagrange multipliers $\lambda_i^{\mathop{\rm opt}\nolimits} \geq 0$, $\mu^{\mathop{\rm opt}\nolimits} \geq 0$, and $\nu_k^{\mathop{\rm opt}\nolimits} \geq 0$.
Similarly, the optimal solution needs to satisfy the following constraints:
\begin{align}
&\lambda_i^{\mathop{\rm opt}\nolimits} (P_{RS_i}-P_{R_i}^{\mathop{\rm opt}\nolimits})=0, \hfill\\
&P_{R_i}^{\mathop{\rm opt}\nolimits} \leq P_{RS_i}, \hfill\\
&\mu^{\mathop{\rm opt}\nolimits} (P_{RS}-\sum \limits_{i=1}^{M} P_{R_i}^{\mathop{\rm opt}\nolimits})=0, \hfill\\
&\sum \limits_{i=1}^{M} P_{R_i}^{\mathop{\rm opt}\nolimits}\leq P_{RS}, \hfill\\
&\nu_k^{\mathop{\rm opt}\nolimits} ({B_k}-{{ \sum_{i = 1}^M {{P_{R_i}^{\mathop{\rm opt}\nolimits}}{{\left|{{h_{{R_i}M_k}}} \right|}^2}}}})=0, \hfill\\
&{{ \sum_{i = 1}^M {{P_{R_i}^{\mathop{\rm opt}\nolimits}}{{\left|{{h_{{R_i}M_k}}} \right|}^2}}}} \leq {B_k} .\hfill
\end{align}

In this case, we also provide a lemma concerning the intervals containing the optimal multiplier.

\textbf{\emph{Lemma 6:}} The optimal $\lambda_i^{\mathop{\rm opt}\nolimits}$, $\mu^{\mathop{\rm opt}\nolimits}$, and $\nu_k^{\mathop{\rm opt}\nolimits}$
satisfying (53)-(58) are respectively within $[0,\lambda_i^\textrm{max}]$,
$[0,\mu^\textrm{max}]$, and $[0,\nu_k^\textrm{max}]$,
where $\lambda_i^\textrm{max}={C_i}$, $\mu^\textrm{max}=\min_i \{C_i\}$,
and $\nu_k^\textrm{max}=\min_i \Big\{\frac{C_i}{{{\left|{{h_{{R_i}M_k}}} \right|}^2}}\Big\}$.

\item[\emph{$\bullet$}] \textbf{Optimal $P_M$ under fixed $P_{R_i}$}:
With the $P_{R_i}$ fixed, $\mathcal{R}_R^{BF}$
is monotonically decreasing in $P_M$. The optimal $P_M$ is achieved
at
\begin{equation}
\label{PMBF}
\begin{split}
P_M^{{\mathop{\rm opt}\nolimits} }&=\max \{P_k^{{\mathop{\rm cand}\nolimits} }\},\hfill\\
P_k^{{\mathop{\rm cand}\nolimits} }&= \frac{(2^{\mathcal{R}_{MS}}-1)\sum\limits_{i = 1}^M {P_{R_i}}{\left|
{h_{R_i M_k}} \right|}^2}{{\left|
\textbf{h}_{M M_k} \textbf{w}_k \right|}^2 -(2^{\mathcal{R}_{MS}}-1)\sum \limits_{j =
1,j \ne k}^K {{{\left| \textbf{h}_{MM_{k}} \textbf{w}_j
\right|}^2}}}, \hfill\\
&\quad k=1,2,...,K.\hfill\\
\end{split}
\end{equation}
\end{itemize}

The algorithm can be summarized as follows.
\begin{algorithm}[htb]
\caption{The RUEs' sum rate optimization for BF.}
\label{alg:1}
\begin{algorithmic}[1]
\STATE \textbf{Initialize} All primal variables $P_{R_i}$,
and $P_M$.
\REPEAT
\STATE \textbf{Step 3:} Compute the multipliers $\lambda_i$, $\mu$, and $\nu_k$;
\STATE \textbf{Step 4:} Compute the $P_{R_i}$ and $P_M$ according to (\ref{PRIBF}) and (\ref{PMBF});
\STATE \textbf{Step 5:} Update the optimal $P_M^{\mathop{\rm opt}\nolimits}, P_{R_i}^{\mathop{\rm opt}\nolimits}$;
\STATE \textbf{Step 6:} Compute the achievable RUEs' sum rate $\mathcal{R}_R^{BF}$;
\UNTIL Convergence.
\end{algorithmic}
\end{algorithm}

In Algorithm \ref{alg:1}, each step can be done with a closed-form manner
and the value of each variable can be easily calculated, which makes
the proposed algorithm efficiently work. In Step 3, the complexity of computing the multipliers is $\mathcal{O}(MK)$ mainly due to the computation of $\nu_k$. With these three multipliers obtained, the rate computation procedure
in Step 4 requires a computational complexity in the
order of $\mathcal{O}(K^2N_B)$, which mainly depends on the
optimal MBS's transmit power design (\ref{PMBF}). In Step 5,
the additional computational complexity for updating
all the optimal power is $\mathcal{O}(M)$. The last Step 6 of
computing the achievable RUEs' sum rate $\mathcal{R}_R^{BF}$ needs a computational complexity with $\mathcal{O}(MKN_B)$.
Considering a typical network scenario with $N_B > M > K$, the computational complexity of Algorithm \ref{alg:1} per iteration
is $\mathcal{O}(MKN_B)$, which mainly comes from the calculation of the optimal MBS's transmit power (\ref{PMBF}).
Actually, under a proper initialization of $\{P_{R_i},P_M\}$
and the determined step size of multipliers, the number of iterations is not large and the proposed algorithm can quickly converges, which has been demonstrated in the following simulation results.

\section{Numerical Results}

In this section, the performance of collaborative precoding IC and BF algorithms in the physical layer is first evaluated. Then, the IC and BF based CRRA solutions are simulated and discussed. In particular, several performance metrics including the system outage probability, sum capacity, and average BER are presented for the collaborative precoding IC and BF algorithms. The RUEs' aggregated rates under IC and BF are considered to evaluate the proposed CRRA solution's performance. To match well with the concerned system model, it is assumed that an H-CRAN scenario consisting of one MBS with one MUE, and $M$ RRHs with $M$ RUEs is considered. The MBS is
located in the center of the cell area with a radius of $500$ meters, while the RRHs and MUE are uniformly distributed in the coverage area of the MBS. The RUE is uniformly distributed in the coverage area of each accessed RRH with a radius of $50$ meters.

\begin{figure}
\centering\vspace*{0pt}
\includegraphics[width=0.52\textwidth]{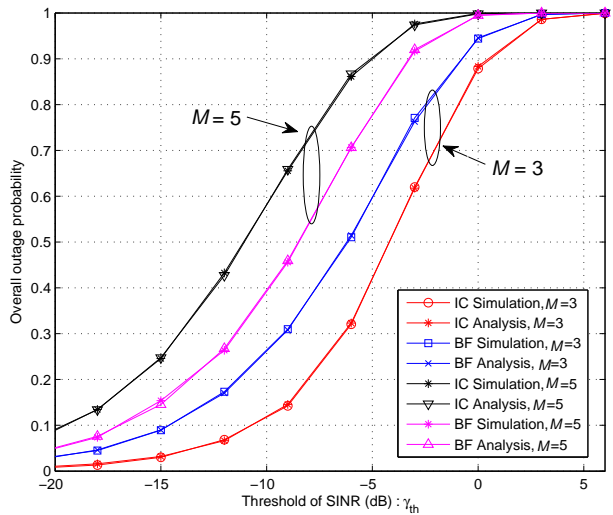}
\setlength{\belowcaptionskip}{-100pt} \vspace*{-10pt}
\caption{Overall outage probability under IC and BF schemes}
\label{out}\vspace{-10pt}
\end{figure}

Fig. \ref{out} shows the system outage probability under different
precoding schemes as functions of the SINR threshold ${\gamma _{th}}$,
and the system outage probability grows of course as the threshold
of SINR increases. The number of antennas on the MBS is set to six,
and one MUE is considered. The Monte Carlo simulation results match
well with those indicated by the presented closed-form overall
outage probability expressions. When $M$ is set to $3$, BF outperforms
IC due to its capability to increase the received signal power
strength. However, when $M$ is $5$, IC is preferred because it can
alleviate the dominating interference. The outage probability gap
for IC between $M = 3$ and $M = 5$ is larger than for BF,
which suggests that IC is more sensitive to the number of RRHs.

\begin{figure}
\centering\vspace*{0pt}
\includegraphics[width=0.50\textwidth]{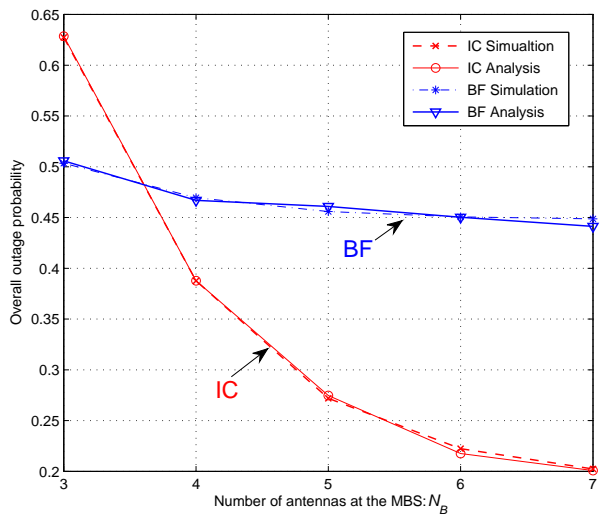}
\setlength{\belowcaptionskip}{-100pt} \vspace*{-10pt}
\caption{Overall outage probability versus the number of antennas
$N_B$} \label{NB}\vspace{-10pt}
\end{figure}

Next, the impact of the number of antennas on the MBS is shown in Fig. \ref{NB}, where we set $M
= 2$, $\gamma_{th} = 0$ dB and one MUE is considered. The overall outage probability
decreases with an increasing number of antennas on the MBS. When $N_B$ is
relatively large, IC becomes better with the optimization in
(\ref{opt}). When $N_B$ is relatively small, BF outperforms IC. This result demonstrates that a large number of
antennas at the MBS is preferred to increase system reliability when
the number of antennas at the RRHs
is fixed.

\begin{figure}[!h]
\centering \vspace*{0pt}
\includegraphics[scale=0.6]{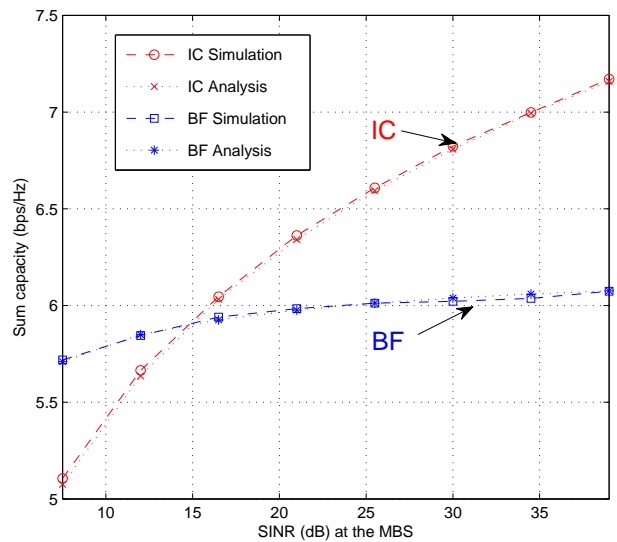}
\setlength{\belowcaptionskip}{-100pt} \vspace*{-10pt}
\caption{System sum capacity under IC and BF schemes versus the SINR (dB) at the MBS for H-CRANs.}
\label{capacity3}
\end{figure}

Fig. \ref{capacity3} shows the system sum capacity under the two
precoding schemes versus SINR at the MBS with $M=2$ and $N_B=6$, where the
system capacity obviously grows as the MBS's SINR increases. The Monte Carlo simulation results match
well with those indicated by the presented system sum capacity expressions. Moreover, the BF scheme
outperforms the IC scheme in the low SINR region due to its
capability of enhancing signal power
strength. However, the IC scheme is preferred at medium to
high SINR because it can alleviate the dominating interference to other MUEs and RUEs.

\begin{figure}[!h]
\centering \vspace*{0pt}
\includegraphics[scale=0.55]{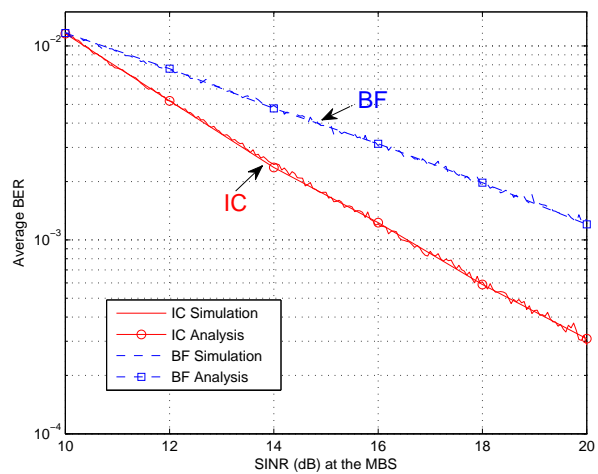}
\setlength{\belowcaptionskip}{-100pt} \vspace*{-10pt}
\caption{Average BER under IC and BF schemes versus the SINR (dB) at the MBS for H-CRANs.}
\label{ber}
\end{figure}

Furthermore, the impact of SINR at the MBS on the average BER is depicted in Fig. \ref{ber}.
The average BER decreases obviously as the SINR of the MBS increases. We can conclude that in the relatively high SINR region, the average BER under IC is lower than that under BF due to the elimination of inter-tier interference from the MBS to the RUEs.

\begin{figure}[!h]
\centering \vspace*{0pt}
\includegraphics[scale=0.65]{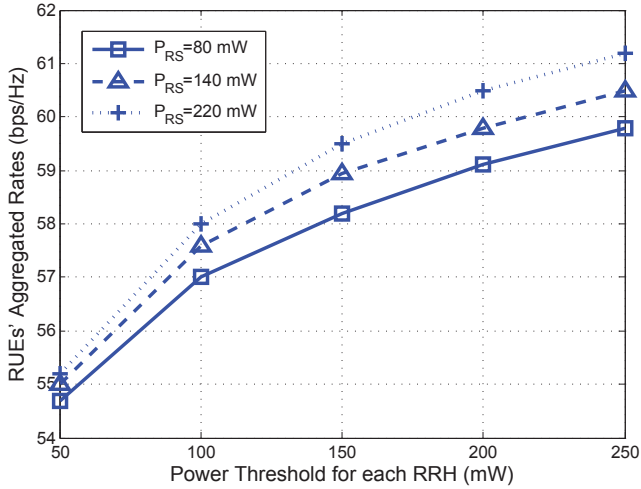}
\setlength{\belowcaptionskip}{-100pt} \vspace*{-10pt}
\caption{RUEs' aggregated rates under the IC scheme versus the power threshold (mW) for H-CRANs.}
\label{optPowerIC}
\end{figure}

\begin{figure}[!h]
\centering \vspace*{0pt}
\includegraphics[scale=0.65]{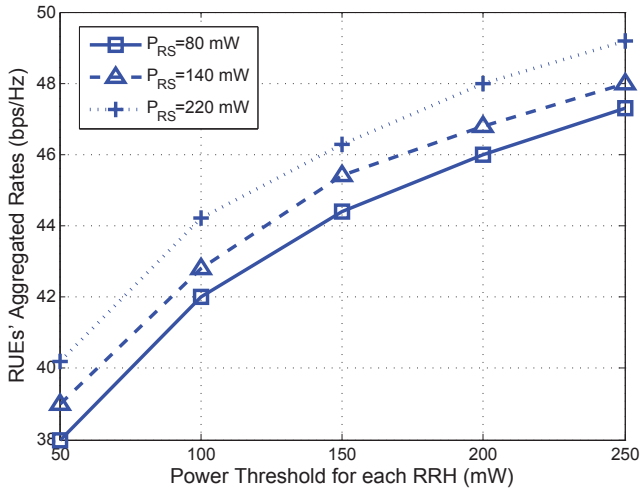}
\setlength{\belowcaptionskip}{-100pt} \vspace*{-10pt}
\caption{RUEs' aggregated rates under the BF scheme versus the power threshold (mW) for H-CRANs.}
\label{optPowerBF}
\end{figure}

Fig. \ref{optPowerIC} and Fig. \ref{optPowerBF} show the RUEs' aggregated rates under the two
precoding schemes versus the power threshold with $M=2, K=3$, $N_B=6$, and $P_{MS}=1000 \text{mW}$.
It is observed that the threshold, as we expect, can increase the RUEs' sum rates.
As the results showed, the RUEs' aggregated rates increases with the power limit of each RRH $P_{RS_i}$ for both IC and BF. This is reasonable since a larger power threshold makes the available power range larger, which leads to larger sum rates. Besides, under the
assumption of the same power limit on each RRH, a larger total power threshold $P_{RS}$ also makes it possible to obtain better performance.

As shown in Fig. 4, it is not clear which precoding scheme outperforms the other one since the dominating factors may change under different SINRs. Fortunately, comparing with Fig. \ref{optPowerIC} and Fig. \ref{optPowerBF} for IC and BF, respectively, the sum rate performance of IC is often better than that of BF under the relatively high SINRs of the MBS because the inter-tier interference has become the biggest challenge impacting the capacity performance when the SINR at the MBS is sufficiently high as shown in Fig. 4. Note that in the regime of low SINRs at the MBS, BF based CRRA often outperforms IC based CRRA because the desired signal strength is low, which dominates the performance of inter-tier interference suppression.

\begin{figure}[!h]
\centering \vspace*{0pt}
\includegraphics[scale=0.65]{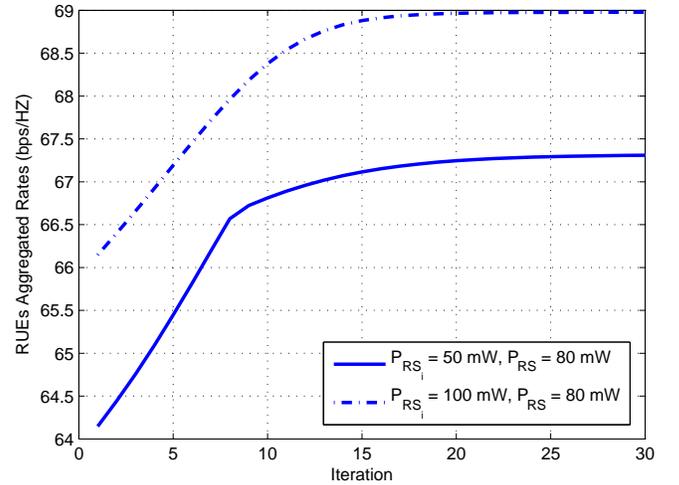}
\setlength{\belowcaptionskip}{-100pt} \vspace*{-10pt}
\caption{Convergence behavior of RUEs' aggregated rates under the IC scheme.}
\label{conv}
\end{figure}

Although a rigorous theoretical proof for the convergence
of the proposed algorithm is not yet available, the RUEs' aggregated rates under IC have been shown in Fig. \ref{conv} to demonstrate the proposal can quickly converge. It is shown that the proposed algorithm can converge with roughly $20-30$ iterations under any $(P_{PS_i},P_{PS})$ set, which indicates that the proposal can work efficiently with low complexity.

\section{Conclusion}

In this paper, we have considered techniques for suppressing the inter-tier interference between the macro base station and the remote radio heads in heterogeneous cloud radio access networks (H-CRANs) in both the physical layer and the upper layer. In particular, the interference collaboration (IC) and beamforming (BF) precoding schemes have been presented to suppress the inter-tier interference in the physical layer, and cooperative radio resource allocation (CRRA) has been optimized in the upper layer. Furthermore, expressions for the overall outage probability, system capacity, and average bit error rate under IC and BF precoding schemes have been derived. Optimal CRRA solutions based on IC and BF have been proposed. Both analytical and simulation results have shown that whether IC or BF provides better performance depends on the H-CRAN configuration, including the number of antennas on the macro base station, the number of remote radio heads, and the target signal-to-interference-plus-noise ratio threshold.

\appendix

\subsection {Proof of Lemma 1}
Given $X \sim \chi _{2L}^2$ and $Y \sim
\chi _{2M}^2$, the CDF of $X$ and the probability density function
(PDF) of $Y$ can be expressed as
\begin{equation}\label{x}
\begin{gathered}
{F_X}(x) = 1 - {e^{ - x}}\sum\limits_{k = 0}^{L - 1}
{\frac{{{x^k}}}{{k!}}}, \hfill\\
{f_Y}(y) = {e^{ - y}}\frac{{{y^{M - 1}}}}{{(M - 1)!}},\hfill\\
\end{gathered}
\end{equation}
respectively. By defining $Z \buildrel \Delta \over = \frac{X}{aY +
b}$, its CDF can be expressed as
\begin{equation}\label{1}
\begin{gathered}
{F_Z}(z) =\int_0^\infty {{F_X}(ayz + bz)} {f_Y}(y)dy \hfill\\
\quad\quad\quad = 1 - \frac{{{e^{ - bz}}}}{{(M -
1)!}}\sum\limits_{k = 0}^{L - 1} {\frac{{{{(az)}^k}}}{{k!}}}
\sum\limits_{i = 0}^k {C_k^i} {\left(\frac{b}{a}\right)^i} \hfill\\
\quad\quad\quad\quad{(az + 1)^{- (k + M - i)}}\Gamma (k + M - i).\hfill\\
\end{gathered}
\end{equation}

\subsection {Proof of Lemma 2}

Consider three RVs $X \sim \chi _{2L}^2, Y_1 \sim \chi
_{2M}^2$, and $Y_2 \sim \chi _{2N}^2$, and define $U \buildrel
\Delta \over = a{Y_1},V \buildrel \Delta \over = b{Y_2}$. Then the PDFs
of $U$ and $V$ are given by
\begin{equation}\label{UV}
\begin{gathered}
{f_U}(u) = \frac{1}{{{a^M}}}{e^{ - \frac{u}{a}}}\frac{{{{u}^{M - 1}}}}{{(M - 1)!}},
{f_V}(v) = \frac{1}{{{b^N}}}{e^{ - \frac{v}{b}}}\frac{{{{v}^{N - 1}}}}{{(N - 1)!}},\hfill\\
\end{gathered}
\end{equation}
respectively. By defining $Y \buildrel \Delta \over = U + V$, its
PDF is obtained as
\begin{equation}\label{Yy1}
\begin{gathered}
{f_Y}(y) = \frac{1}{{\Gamma (M)\Gamma (N){a^{M}}{b^{N}}}}{e^{ -
\frac{y}{{b}}}} \hfill\\
\quad\quad\quad \int_0^y {{u^{M - 1}}{{(y - u)}^{N - 1}}{e^{
- (\frac{1}{{a}} - \frac{1}{{b}})u}}} du.\hfill\\
\end{gathered}
\end{equation}

Following 3.383.1 in~\cite{Table},
\begin{equation}
{\int_0^u {{x^{v\! - 1}}(u \!-\! x)} ^{\mu \! - 1}}{e^{\beta x}}dx\!
= \!B(\mu ,v){u^{\mu\! + v \!- 1}}{}_1{F_1}(v;\mu + v;\beta u),
\end{equation}
where
\begin{equation} _1{F_1}(\alpha;\gamma;z)=1+\frac{\alpha}{\gamma}{\frac{z}{1!}}\!+\!\frac{{\alpha}(\alpha+1)}{{\gamma}{(\gamma+1)}}{\frac{z^2}{2!}}
\!+\!\frac{{\alpha}(\alpha\!+\!1)(\alpha\!+\!2)}{{\gamma}{(\gamma\!+\!1)(\gamma\!+\!2)}}{\frac{z^3}{3!}}+...
\end{equation}
is a confluent hypergeometric function. Eq. (\ref{Yy1}) can be rewritten as
\begin{equation}\label{Zz1}
\begin{gathered}
{f_Y}(y) = \frac{1}{{\Gamma (M + N){a^{M}}{b^{N}}}}{e^{ -
\frac{y}{b}}}{y^{M + N - 1}} \hfill\\
\quad\quad\quad{}_1{F_1}\left(M;N + M; -
(\frac{1}{a} - \frac{1}{b})y \right).\hfill\\
\end{gathered}
\end{equation}

By defining $Z=\frac{X}{Y}$, since $X$ and $Y$ are independent, the CDF of $Z$ can be expressed as
\begin{equation}
{F_Z}(z) = \int_0^z {\frac{{{a^M}{b^N}{x^{L - 1}}}}{{\Gamma
(L)\Gamma (M + N)}}I(M,N,L,a,b,x)dx},
\end{equation}
where
\begin{equation}
\begin{gathered}
I(M,N,L,a,b,x) = \int_0^\infty {{x^{M + N + L - 1}}{e^{ - (y +
\frac{1}{b})y}}}\hfill\\
\quad\quad{{}_1{F_1}\left(M;N + M; - (\frac{1}{a} - \frac{1}{b})y \right)} dy.\hfill\\
\end{gathered}
\end{equation}

To obtain a closed-form CDF expression, $Z$ can be approximated as
$Z \approx \frac{X}{{a{Y_1} + bN}}$. Then according to Lemma
1, the CDF of $Z$ can be approximately expressed as
\begin{equation}
\begin{gathered}
{F_Z}(z) \cong 1 - \frac{{{e^{ - (K-1)z}}}}{{(M - 1)!}}\sum\limits_{k = 0}^{N_B - 1} {\frac{{{{(az)}^k}}}{{k!}}} \sum\limits_{i = 0}^k {C_k^i} {\left(\frac{bN}{a} \right)^i} \hfill\\
\quad\quad\quad\quad {(az + 1)^{ - (k + M - i)}}\Gamma (k + M- i). \hfill\\
\end{gathered}
\end{equation}

\end{document}